# The electromagnetic "memory" of a dc-conducting resistor: a relativity argument and the electrical circuits

*Emanuel Gluskin*

Galilean Sea Academic College, Ort Braude Academic College (Carmiel). Holon Institute of Technology, Electrical Engineering Department of the Ben-Gurion University, Beer-Sheva 84105, Israel. gluskin@ee.bgu.ac.il

**Abstract**: A circuit-field problem is considered. A resistor conducting a constant current is argued to be associated with electromagnetic energy accumulated in the surrounded space, though contrary to the case of an inductor or a capacitor, this energy is always associated with *both* magnetic and electrical fields. The circuit-theory point of view saying that a resistor has no electromagnetic memory is accepted, but the necessarily involved (in view of the field argument) capacitance and inductiveness are argued then also not be associated with any memory. The mutually completing circuit and physical arguments are presented as a dialog between a physicist and an electrical engineer. *How can you call "parasitic" the elements that represent the fields due to which your resistor at all receives the energy?!* -- asks the physicist finally.

## 1. Introduction

It is both a point of principle and a quite practical and pedagogical item to well understand the physical foundations of circuit theory, at least in the basic scope in which Kirchhoff's equations appear from Maxwell's equations. The present work can be seen as related to this scope, though no bulky equations or expressions associated with any PDE-s appear here, i.e. the discussion is a qualitative one, starting from an argument related to the theory of relativity, and having the whole stress on the logical side. We focus on the interesting aspect of energy accumulated in the space around resistor with electrical current flowing through it. The unusual slant is in "equalizing the rights" of the circuit theory and the physical outlooks that are presented as completing each other.

   While considering the circuit physics basics, we do not completely forget what for the electrical circuits are needed at all, which is expressed in an unusual for such a discussion involvement of the concept of memory. The memory can be associated with the energy accumulated in the electrical and magnetic fields in (around) the circuit.

   Making the meaning of the words clear, we reject the common among circuits specialists opinion that a resistor does not accumulate electromagnetic energy, while the opinion that a resistor does not have electromagnetic memory is supported. As a whole, the presented work completes circuit theory textbooks that often over-simplify the physical situation regarding real circuits, and the material should not be missed by the teachers of electrical engineering, the students, and engineers.

   The given argumentation agrees, of course, with the known [1-9] frequency criterion for transfer from lumped to distributed model for a circuit of a given dimension (duration). Calculations of the radiation parameters and time constants can be found in these references. However, the transfer of the initially variable fields to those constant (nonzero) taking place in the final dc state, and, especially, the connection with the system-theory concept of memory are nontrivial points.

Using below the term "heating" as regards the resistor, we just mean transfer the energy from the battery to the resistor. The conditions of cooling the resistor, which also influence its temperature, as well as this very temperature, do not interest us here.

## 2. The combined circuit theory and physics outlook on the conductivity problem

From the circuit-theory point of view, resistor is an element that realizes the voltage-current characteristic $v = Ri$ (more generally $v = f(i)$ with a unique, possibly nonlinear, function $f(.)$, and only with this feature the resistor is included in circuit theory that is, essentially, a mathematical discipline, -- so why should one think about physics here? There is, however, a good reason for the latter, which is never well formulated in the circuit theory, -- ***though the concepts of voltage drop v on an element and of electrical current i flowing via this element are clearly defined, -- these parameters/variables cannot relate only to the conducting resistor. The very existence of v and i indicates the presence of (necessity in) some other elements, and the reasonability of inclusion of these elements into the circuit's scheme should be carefully considered.***

The above statement is, first of all, a physical one since the relation of the other elements to the resistor is associated with the physical fields. That these fields are relevant to discussion of a "resistive" circuit immediately follows from the fact, -- well-known from the theory of relativity, -- that no signal can propagate quicker than light, i.e., an electromagnetic wave.

Indeed, if only algebraic equations (those describing the resistors) are involved in the analysis, then the problem arises regarding the fact that even a very large connection of numerous resisters would *immediately* transfer the influence of the battery (the "signal") to a distance, and an effective heating of any however distanced resistor of the connection would start immediately. This is especially simply seen for any connection of *linear* resistors, when the voltage drop on any resistor mathematically comes out to be directly proportional to the battery voltage $E$. If at the input of such a network there is voltage $E$, then for a certain distant resistor $R$ we have for its voltage drop $v_R \sim E$ and for its power $P_R \sim v_R^2 \sim E^2$. With connection of the battery, $v_R$ immediately jumps from the zero to this value *which means an infinite velocity of a signal informing us that the battery is connected.* (We have the same physical time at the input and the "output".) Similarly, $E \to 2E$ *immediately* causes $v_R \to 2v_R$ and $P_R \to 4E^2$, etc. .

It becomes obvious, in view of the relativity argument, that a wave including electrical and magnetic field must participate the real process(es) occurring after closing the switch that connects the battery, and the heating can start after this wave reaches a resistor, not earlier. If so, this wave should be bringing the energy of the heating.

One notes, furthermore, that the heating should not disappear immediately with disconnecting of the battery. Indeed, if the voltage on the distanced resistance would immediately disappear this would mean a signal informing about the disconnection, i.e. we have absolutely the same causality problem as with connecting the battery.

Of course, by itself, such argument relates to any lumped, not necessarily "algebraic" (resistive), linear circuit, since consideration of such circuit leads in circuit theory, via application of Kirchhoff's equations, to the linear integral input-output relation

$$f_{output}(t) = \int_{-\infty}^{t} h(t,\lambda) f_{input}(\lambda) d\lambda \qquad (1)$$

where, for the case of constant elements, $h(t,\lambda) = h(t-\lambda)$, and if, e.g., at some moment $t$, $f_{input} \to 2 f_{input}$ (i.e. $f_{input}(t^+) = 2 f_{input}(t^-)$), then some change in $f_{output}(t)$ also immediately *starts* to occur with connecting the battery. However, if the circuit is not purely algebraic, this change is immediate in $df_{out}/dt$, not in $f_{output}(t)$, which should be less suitable for measurement.

In any case, it is obvious that, having in (1) $f_{output}$ as a function of only time and not any spatial variable, we cannot observe any delay, i.e. have a finite velocity of the signal.

Thus, the relativity argument makes the fields be the central object of the process, while the resistor becomes a kind of boundary for the fields.

Since the electrical and magnetic fields involved are associated, from the circuit point of view, with some capacitances and inductivities which are the "other elements" mentioned in the above statement, a physicist categorically rejects seeing the real battery-resistor (*E-R*) connection as a circuit including only *E* and *R* elements. However, when physicist teaches technical students physics basics, he should know what will be the positions of the teachers of the technical courses given to the same students, and try to think out the circuit-theory stresses.

Using below the circuit theory term "memory", we introduce something new in the classical physical argumentation, and the logic of the situation (and not any calculations based on Pointing's vector) is the focus. The purely physical calculations are well described in the references given (and are not the concern here), but this logic should not be missed in the education of the modern electrical engineering students.

The discussion is centered around the facts that *there is* some energy, both electrical and magnetic, *in the dc state* of the *E-R* circuit, but the *memory* associated with the existence of the energy is too partial for one to say that the resistor possesses electromagnetic memory. Accumulation of energy is always associated with elements (specially included as lumped elements, or the "stray" ones) described by differential equations whose solutions generally depend on initial conditions, which is some memory. However, the situation is twofold, since resistor both requires the field elements to be introduced and makes the situation in the extended circuit not a dynamic one. The latter means that one cannot know, in the established dc state, what were the initial conditions for the states of the field elements. The dissipation of energy in the resistor, and the "dissipation of memory" in the associated field elements cannot be separated.

Let us consider the following imaginary dialog between an electrical engineer (**EE**) and a physicist (**P**) showing the gap between the approaches, and their mutual completion, and attempting in "sewing" the different points of view.

### 3. The traditional circuit-theory argument and the physical reality

**EE:** The concept of memory is very important in modern electronics systems. For a causal system, -- and only such systems should interest physicists, -- memory means the dependence of the state of a system on the initial conditions related to the past. When speaking only of simple elements, electrical dynamical systems are composed of inductors, capacitors and resistors. Inductors and capacitors have electromagnetic "memories", but resistors do not. However, the resistors can prevent the inductors and capacitors from remembering their past, and thus "kill" their memory. Thus, a dynamic system, i.e. one that can have memory of its past, does not always really have memory. It is worth stressing that a circuit can be well understood if one uses the concept of memory.

**P**: You speak about mutual influence of the elements, using some information or operation terms. I do not know what you mean, but dislike already your first insistent saying that resistor has no electromagnetic memory.

**EE**: *Of course, resistor has not any memory*. Let us compare it with inductor and capacitor. The respective physical laws of these elements are, $v = Ldi/dt$ and $i = Cdv/dt$, including *time-derivatives*, and thus Kirchhoff's equations for circuits including such elements are *differential equations* whose general solutions include constants to be found from the initial conditions. This connection of the process/state at $t > 0$ with the initial state at $t = 0$ obviously means a memory.

Another expression of the memory is given by the accumulated energy. The inductor and the capacitor accumulate, respectively, the energies $W_L = Li^2/2$ and $W_C = Cv^2/2$, and since the process of energy accumulation depends on the processes $v(t)$ and $i(t)$ during the time period passed, -- the accumulated energy too means a memory, even if in a much more partial sense, not a dynamic one.

Contrary to that, no influence of any initial conditions is expressed in the algebraic relation $v = Ri$, and where is there an accumulated electromagnetic energy that might be taken back from the resistor as one can take back the energy accumulated in a capacitor or an inductor? A resistor just generates thermal losses. In terms of analytical mechanics, the instantaneous *power* of the resistor $p_R = vi = Ri^2$ is not any "full differential" (while the powers $(Ldi/dt)i \sim d(i^2)/dt$ and $vCdv/dt \sim d(v^2)/dt$ are such) and thus we have no integral of movement, i.e. no electrical energy stored.

**P**: You want to understand all in terms of lumped elements circuits. Though for me as a physicist this is, finally, unsatisfactorily, let me try to say something in these terms. For a series *L-R* connection, we have, using the expressions you used,

$$p_R = Ri^2 \equiv (2R/L)(Li^2/2) \sim W_L,$$

i.e. there is immediate connection, in the circuit, of $p_R$ with the magnetic energy $W_L$ of the inductor. For an oscillatory *R-L-C* circuit we have, as you can easily check,

$$p_R = - d(Li^2/2 + Cv^2/2)/dt \qquad (2)$$

i.e. *$p_R$ can be* presented as the full differential of some *electromagnetic* energy, including both $i$ and $v$, and associated with a memory.

**EE**.  If you speak about a series *RLC* circuit, then '*i*' relates to the resistor, but '*v*' in (2) relates only to the capacitor.  For a parallel circuit, '*v*' would relate to the resistor, but '*i*' only to the inductor.

**P**.  I shall show you that these *v* and *i* are associated in our circuit with the resistor.

**EE**.  Where are these *L* and *C* elements in the simple *E-R* circuit, including only battery *E* and the resistor, which all EE students learn first, before they know anything about transients in first-order, *LR* or *CR*, circuits, or the oscillatory processes in *RLC* circuits?  Ohm's law is always studied for a pure resistor directly connected to a battery!

**P**.  The physical situation certainly allows us to introduce some *C* and *L* elements, and in my explanation, which follows, I shall use only some very basic electrodynamics concepts that you can find, e.g., in [1,2]. Let us look at $v = Ri$  (or any  $v = f(i)$  that would be relevant to a nonlinear resistor) *as physicists*. ***The very fact of the presence of voltage and current means the presence of electric and magnetic fields, even in the dc state, which you are ignoring***.  Indeed, according to Ampere's law, the current necessarily causes a magnetic field **H** around the resistor, and according to the boundary condition of continuity of the tangential component of the electrical field **E** that is associated with voltage *v*, there is some electrical field around the conductor. You have to see that **E** and **H** are responsible for the heating of the resistor. Figure 1 shows that close to the surface of the resistor the Poynting's vector of these fields, **S** = **E**×**H**, is oriented towards the resistor and, e.g. for the simple cylindrical form of the resistor, the flow of the energy through the surface $2\pi rd$ is

$$p = S 2\pi rd = EH\, 2\pi rd = [(v/d)\, i/(2\pi r)]\, 2\pi rd = vi = p_R. \qquad (3)$$

For a linear resistor, $v = Ri$,  $p_R = vi = ri^2$, but the above derivation holds also for any nonlinear resistor, since no certain $v = f(i)$ is used in it.  That for $p_R$ nonzero both **E** and **H** must be nonzero is made obvious.

Furthermore, as any flow, the flow of electromagnetic energy, i.e. Poynting's vector, is the density of the flowing substance multiplied by its velocity, and Fig. 1 also schematically shows that the energy propagates from the source via the surrounding space and enters the resistor.  In terms of the energy densities [1,2], this means that

$$\mathbf{S} = \mathbf{n} c_m (w_E + w_H) = \mathbf{n} c_m (\varepsilon E^2 \backslash 2 + \mu H^2 \backslash 2)\,, \qquad (4)$$

where the unit vector **n** is in the direction of the flow, and $c_m = c(\varepsilon\mu)^{-1/2}$ is the light-velocity in the medium. Comparing (4) with (3), we see that  $p_R \sim (w_E + w_H)$.

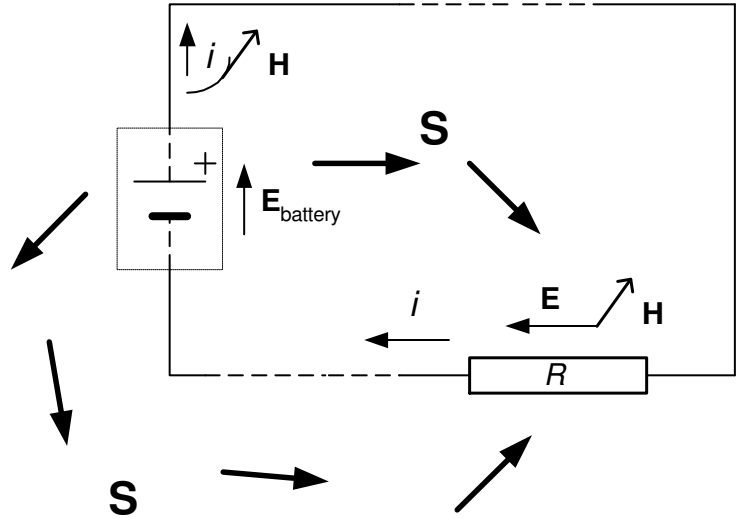

Fig. 1: The energy (power) enters the resistor from outside because of the fields **E** and **H**, and thus $p = p_R = Ri^2$ (or, more generally, $vi$) is obtained. The length of the resistor is $d$, and the radius $r$; thus the relevant part of the surface $A$ of the resistor is $2\pi r d$, which is used in (2) The d.c. state is in the focus.

The next step is to note that the complete electromagnetic energy W, accumulated in the whole space, is the integral of $w_E + w_H$ over the volume:

$$W = \iiint (w_E + w_H)\, dV,$$

and it follows from $p_R \sim w_E + w_H$ that

$$p_R \sim W,$$

as in equation (2) that you criticized. (Consider that in (2) the total energy of the capacitor and the inductor is weakly exponentially damping, i.e. the derivative is proportional to the function.)

It is worth noting that the connection between $S = p_R$ and W also directly follows from the fact that in the dc state the resistor is the whole (conductive) boundary for the fields.

*Thus, first of all, we cannot speak about an "E-R circuit" from the very moment of closing the switch; one can do this after the initial transient associated with the field processes will be ended, while the latter depends on the circuit's whole physical structure. This is ignored in the textbooks; the description always starts from t = 0, without any reservations, which, from the physical point of view, is serious miss, even though this transient is usually very quick* **[6,7,9]**.

*Moreover, you cannot ignore the fields even in the dc state, and, in principle, the L and C elements remain with the energy accumulated by them. That in the dc state you do not feel these elements does not mean that the fields are not important also in this state. Of course, you will feel the fields and the additional elements each time when a change of E will cause a new transient, but these field are responsible for heating the resistor in any case.*

Regarding the circuit model, I think that some chain-type *LC* model circuits could well represent the system after closing the switch, when a quick transient occurs and we obviously have a distributed system. When after some time the steady situation is established, the model remains, with all of the capacitors in it being finally charged.

To complete the physical description of the circuit, I should recall that *inside* the resistor with the flowing current there also is some magnetic field. However, there is nothing qualitatively new in this for my point after the external magnetic field is already taken into consideration.

In any case, the power consumed by a resistor in a "d.c. process" *can be* associated with some electromagnetic energy accumulated in the system, and thus the conducting resistor has some "memory".

**EE**: I see your point, and regarding the *initial transient*, I can even help you, suggesting the following argument in circuit terms. Before the switch that connects *E* and *R* is closed, the current in the circuit was zero. When ignoring the non-lumped *L* and *C* elements that, as you argue, are present, we have, upon closing the switch, the momentary jump of *i*(*t*) from the zero value to the value *E*/*R*. Such a jumpy *time-function i*(*t*) has an infinite frequency spectrum, including the high frequencies which are associated, in terms of electromagnetic waves, with some *short* wavelengths that do not satisfy the known [3,4] conditions of a lumped circuit. This necessarily means that electromagnetic waves are radiated out of the circuit, -- just like the radiation which makes many industrial switched circuits "noisy", -- and together with the radiation of the energy out of the circuit, the energy also starts to enter the resistor.

Thus, I could predict that an electromagnetic field should appear at the initial stage, and that during the initial transient the circuit was not *E-R*, but *E-R-L-C*, and the same argument remains when you then abruptly change the battery voltage from *E* to any other value.

However, I confess that the showing by you that the electromagnetic energy inevitably exists also in the d.c. state *when nothing is changed*, is unexpected for me. Nevertheless, ***I am sure that any resistor has no electromagnetic "memory"***.

**P**. The accumulated electromagnetic energy reminds us about the initial transient leading to the steady state. Isn't this a memory?

**EE**. In a too partial or vague sense. For me, the real memory associated with a circuit is a dynamic one. The initial conditions should be influencing, and it should be possible to *use* the energy accumulated in the space. It is *easy* to use the accumulated, *localized* in a lumped capacitor, energy. The capacitor can be discharged via some load, which is very common in power electronics where periodical charging and discharging the capacitor define the average current and average power of the load.

**P**. However, you can open in the dc circuit the switch that connects the resistor to the battery, interrupting the current, and get a *pulse of radiation*, which (for *R* known) can even be used as an indication (measurement) of the level of charge of the battery. This is unpractical, of course, but, in principle, seems to me to be not so far from any use of $Li^2/2$ or $Cv^2/2$ when you have lumped elements, even though **E** or **H** are not separated and we use **S** here.

**EE**: You have to see that system theory teaches us to think in terms of simple "state-variables" [3,4]. The "memory" that you try to ascribe to the resistor is directly associated with the specific combination **E**×**H** of the fields, which is finally equivalent to the specific combination $vi = p_R$ of the usual state-variables $v$ and $i$. That, contrary to $v$ and $i$, **S** and $p_R$ themselves are *not* any standard state-variables, is a problem for me, and it is difficult for me to constructively connect the physical picture you are drawing with the system concepts to which I am used and in which I have to think.

**P**: Well, you are not ready to see the circuit just as a physical system, and your stress on the (merely informational) aspect of memory is not close to me. It is obviously not easy to "sew" the physics and the system points of view. Nevertheless, when I teach technical students, I must consider what they know, in order to well connect the things in the students' heads. Thus I do have to think out all this.

However, it is interesting for me now to better understand your words at the start that a resistor can "kill" the memory of $L$ and $C$ elements. Could you explain this in detail?

**EE**: To explain this, it is sufficient to consider the equation for an *L-R-E* circuit:

$$L di/dt + Ri = E$$

from which

$$i(t) = [i_o - E/R] \exp\{-t/\tau\} + E/R, \qquad (5)$$

where $i_o = i(0)$. It is important that the asymptotic term $E/R$, as well as the time-constant $\tau = L/R$, exists only for a nonzero $R$. For $R = 0$, we would have:

$$L di/dt = E$$

and the linearly increasing $i(t)$:

$$i(t) = i_o + (E/L) t. \qquad (6)$$

Even in (6) the situation with the memory about $i_o$ is not quite simple since the memory tends to zero as time passes, because the ratio of the first term to the second one is $\sim 1/t$, i.e. it becomes *more and more difficult to measure* the constant term $i_o$ in $i(t)$. However in (5) the situation re memory is really dramatic, -- the relative weight of the term $i_o \exp\{-t/\tau\}$ very quickly decays at $t \gg \tau$, and this is done by the nonzero $R$.

Because of the resistor, the inductor very soon forgets its initial current forever. This is what I mean by saying that $R$ "kills" the dynamic memory in $L$. Some "killed memory" should be feature of any dissipative system, caused by the energy consuming elements.

All this is also relevant to your *RLC* model of the *E-R* circuit. In the d.c. state, your $L$ and $C$ elements are without memory in the dynamic sense, since this state is not influenced by the initial conditions. The whole d.c. circuit, including your "field" additions, does not possess any real memory, and *thus* I forgot about the fields and the associated distributed capacitor and inductor which are rather meaningless from the operation point of view. However, I agree that when the current changes in time, these parasitic elements become important.

**P**: *How can you call "parasitic" the elements that represent the fields due to which your resistor at all receives the energy?!*

**EE**: I am thus taught, but, maybe, you can teach the young generation of electrical engineers more appropriately to their actual knowledge. Consider that there is no problem for the students to find material about circuit radiation. To read/study this material, or not to read/study, -- *this is the question*. I would suggest you to at least try to involve the topic of memory in your considerations with the students. It will not take a lot of additional time, and the connection with physics is sufficiently interesting.

### 4.  Summary and final comments and suggestions

Let us conclude the discussion by the following points, some of which are formal conclusions and some (4-8) suggest additional considerations to further think the things out and consolidate one's understanding.

1. The physical background of the simplest electrical *E-R* circuit is not trivial; it includes several heuristically important points. In particular, theory of relativity prohibits having a purely resistive network, because an algebraic map $i \rightarrow v$ leads to an infinite velocity propagation of signals in the system.

2. The statement that resistor is not associated with electromagnetic energy is wrong. However when, because of the resistor, the d.c. state is finally established, the current and voltage functions do not depend on the initial conditions, and despite the presence of some equivalent dynamic elements associated with the accumulated electromagnetic energy, there is no dynamic (associated with the initial conditions) "memory" in the system, just some "memory" about the whole initially occurring transient process.

3. The resistor can be considered as a part of a *boundary* for the electromagnetic field; a *conductive* boundary, and *in the d.c. state, the resistor is the whole boundary*, because the dc circuit does not act as antenna. The voltage drop on the resistor and the conditions of the continuity of the tangential component of *E*, together with the magnetic field caused by the current, direct the Poynting's vector of the field so that the needed power, *vi* (which is $Ri^2$ for a *linear* resistor), enters the resistor.

4. Draw the vector field of the Poynting's vector *in the whole space* (i.e. complete Fig. 1), starting from the battery.

5. Consider whether or not it is possible that in the d.c. state the amount of electrostatic energy accumulated in the space (*not* inside any battery, of course) precisely equals the amount of accumulated magnetic static energy.

6. Consider attempts at "blocking" the fields by using a screen placed between the source and the resistor, for the cases of grounded and ungrounded screen.

7. Consider, in the context of the fields, the resistor to be a *lamp*, either incandescent or (then it is a nonlinear resistor) a fluorescent one [11].

8. Compare the "relativity argument" of Section 2, based on the limitation of velocity of propagation of signals, with such argument used in the known proof that no absolutely rigid bodies exist (otherwise an absolutely rigid rode would transfer the "signal" about its mechanical movement with an infinite velocity). In both cases, the relativity argument forces us to think about the *internal structure* of the physical object. The connections between the atoms in any real rode are electrical. Will an electromagnetic signal appear and come to the object purposed to be moved by the rode, before the atoms of the distanced (close to the object) edge of the rode start to move?

Can such a signal be "composed" not (or not only) of photons, but (also) of some other particles with a zero mass of rest? (Consider Section 11 of [2].)   Physically a very long rod should be very heavy, and the blow (force) that starts to move is, must be huge; thus one *can* think about high energy-processes at the beginning of the rode.

**Acknowledgments**